\documentstyle{article}
\begin{document}

\title{{\bf Managing Null Entries in Pairwise Comparisons}}
\author{W.W. Koczkodaj\thanks{
partially supported by the Natural Science and Engineering Council of Canada}
\\
Computer Science, Laurentian University \\
Sudbury, Ontario P3E 2C6, Canada, icci@nickel.laurentian.ca \and Michael W.
Herman \\
Computer Science, Laurentian University \\
Sudbury, Ontario P3E 2C6, Canada, mwherman@nickel.laurentian.ca \and Marian
Or\l owski\thanks{%
partially supported by the Australian Research Council} \\
School of Information Systems, Queensland University of Technology \\
Brisbane, Q4001, Australia, Orlowski@fit.qut.edu.au}
\date{Sudbury in Canada and Brisbane in Australia, \today}
\maketitle

\begin{abstract}
This paper shows how to manage null entries in pairwise comparisons
matrices. Although assessments can be imprecise, since subjective criteria
are involved, the classical pairwise comparisons theory expects all of them
to be available. In practice, some experts may not be able (or available) to
provide all assessments. Therefore managing null entries is a necessary
extension of the pairwise comparisons method. It is shown that certain null
entries can be recovered on the basis of the {\em transitivity property}
which each pairwise comparisons matrix is expected to satisfy.\newline

\noindent KEYWORDS: {\em uncertainty, incomplete information, pairwise
comparisons, consistency-driven approach, comparative assessments.}
\end{abstract}


\section{Introduction and Basic Concepts}

The pairwise comparisons method, introduced in embryonic form by Fechner (%
\cite{Fech1860}) in 1860, was formalized and extended considerably by
Thurstone (\cite{Thur27}) in 1927. Saaty (\cite{Saaty77}) transformed the
method into a useful tool in 1977 with the addition of hierarchical
structures (previously the $O(n^{2})$ complexity had been a problem for
large $n$). The introduction of triad inconsistency in \cite{Kocz93} within
the framework of a consistency-driven approach (presented in \cite
{KoczMack97,Kocz97,JanKocz98}) was a further refinement that enables experts
to locate and reconsider their most inconsistent assessments.

The consistency-driven pairwise comparisons method stresses that nothing can
replace the assessments of experts and that monitoring of inconsistencies is
only used to point out the most inconsistent judgments to the experts. In
most applications this procedure suffices for a reconsideration and possible
revision of the assessments. However, there may be special circumstances
(e.g. lack of time, or departure of an expert) where a reassessment is not
an option and then the completion of the project depends on the missing
values (i.e. null values) being filled in. This mechanical replacement of
null values (using statistical averages or other heuristics) should only be
done if absolutely necessary. Managing null values can also extend the
application of the consistency-driven pairwise comparisons method to
situations where experts lack sufficient information to form complete
assessments.

In most former contributions involving the pairwise comparisons method, a
pairwise comparisons matrix is defined as an $n$ by $n$ matrix $A=[a_{ij}]$
such that $a_{ij}>0$ for all $i,j=1,\ldots , n$. It will be shown in this
paper that under certain conditions ``for all'' can be replaced by ``for
some'' while the remaining elements may be undefined.

\begin{equation}
A=\left| 
\begin{array}{cccc}
1 & a_{12} & \cdots & a_{1n} \\ 
\frac{1}{a_{12}} & 1 & \cdots & a_{2n} \\ 
\vdots & \vdots & \vdots & \vdots \\ 
\frac{1}{a_{1n}} & \frac{1}{a_{2n}} & \cdots & 1
\end{array}
\right|
\end{equation}

\noindent In general (i.e., when defined), $a_{ij}$ represents an expert's
relative assessment of stimulus (or criterion) $s_{i}$ over stimulus $s_{j}$%
. If the stimuli $s_{i}$ are represented by weights (positive real numbers) $%
w_{i}$, then $a_{ij}=\frac{w_{i}}{w_{j}}$. Since it is normally expected
that $a_{ji}=\frac{w_{j}}{w_{i}}=\frac{1}{\frac{w_{i}}{w_{j}}}=\frac{1}{%
a_{ij}}$, the matrix $A$ is usually assumed to be reciprocal$.$ However,
even this condition can be dropped since non-reciprocal matrices can be
transformed into reciprocal matrices as shown in \cite{KoczOrl98}.

A pairwise comparison matrix $A$ is called {\em consistent} if $a_{ij}\cdot
a_{jk}=a_{ik}$ for $i,j,k=1,\ldots ,n$. While every consistent matrix is
reciprocal (since $a_{ij} \cdot a_{ji}=a_{ii}=1$), the converse is false in
general. Consistent matrices correspond to the ideal situation in which the
exact weights $w_{1},\ldots ,w_{n}$ for the stimuli are available since the
matrix of quotients $a_{ij}=w_{i}/w_{j}$ then satisfies the consistency
condition. Conversely, for every $n\times n$ consistent matrix $A=[a_{ij}]$
there exist positive real numbers $w_{1},\ldots w_{n}$ such that $%
a_{ij}=w_{i}/w_{j}$ (see \cite{Saaty77} for details). The vector $%
w=[w_{1},\ldots w_{n}]$ is unique up to a multiplicative constant and
represents the weights for the stimuli $s_{i}$, $i=1,\ldots ,n$. The
challenge to the pairwise comparisons method comes from the lack of
consistency of the pairwise comparisons matrices that arise in practice. For
an $n\times n$ matrix $A$ which is inconsistent, a consistent matrix $%
A^{\prime }=[w_{i}/w_{j}]$ can be found (see, e.g., \cite{KoczOrl97}) which
differs from $A$ ``as little as possible'' for certain metrics (e.g., the
Euclidean metric in the case of the least squares method). However, the
matrix $A^{\prime }$ may have many elements that are different from the
original matrix $A$ even though in practice an expert may be unsure of only
a few assessments. Thus the location of the most inconsistent assessments is
more important than a precisely computed consistent matrix. However, the
eigenvalue inconsistency index, based on a global property of a matrix,
cannot identify the most inconsistent triad of assessments. This reasoning
resulted in the introduction in \cite{Kocz93} of a finer-grained measure of
inconsistency which permits the location of the most inconsistent triad and
consequently allows a reconsideration of the assessments involved. Moreover,
a formal proof of the convergence of a class of algorithms for reducing the
triad inconsistency is provided in \cite{HolKocz96} while a similar proof
for the eigenvalue inconsistency index does not exist and may not be easy to
provide since the relationship between eigenvalues (a global characteristic
of a matrix) and a process of local improvements of triads is unknown and
rather challenging to examine.

In this paper we propose a further generalization of pairwise comparisons
matrices to situations where one or more assessments are unavailable. We
thus postulate that the pairwise comparisons matrices may contain null
entries. The constructiveness of our approach derives from the observation
that we should be able to process what we have, without expectations (or
assumptions) that the entire matrix is given. In other words, we fill in the
entries in the matrix corresponding to whatever experts assess; hence the
necessity of managing null entries. We note that the representation of
matrices with null entries is easily handled by several array-based
languages (APL2, J, Nial) where arrays are defined as rectangular
collections of items each of which may be another array and in particular an
empty array (which corresponds to our null entry).

\section{Null Entries in a Pairwise Comparisons Matrix}

An arbitrary scale $[1,N]$ (and hence $[1/N,1]$ for the inverses) is used to
compare all stimuli in pairs. Here $1$ stands for equal preference and $N$
for the greatest preference of one stimulus over another. While Saaty uses $%
N=9$, the value $N=5$ works better in practice (\cite{Kocz93}) since a wider
scale only serves to confuse users rather than introduce a greater precision
into a situation where rough subjective assessments are involved. (According
to \cite{Zah93}, all reasonable scales are equivalent for a small enough
inconsistency). The classical pairwise comparisons method assumes that
comparisons for all possible combinations of stimuli are available. In
practice there are many situations where comparing certain stimuli may be
difficult. An ad hoc solution to uncertainty is using a value of {\em one}
for both {\em equal importance} and {\em unknown importance}. But the
unknown value is {\em any value} and using $1$ in this situation is
reflected in Henry Ford's reference to the color assortment of his famous
model T: ''you can have it in any color you wish provided it is black''. In
fact substituting $1$ for an unknown value can easily lead to incorrect
conclusions as is shown in the following example

\begin{equation}
\left| 
\begin{array}{ccc}
1 & 1 & 5 \\ 
1 & 1 & {\bf 1} \\ 
\frac{1}{5} & {\bf 1} & 1
\end{array}
\right|
\end{equation}

\noindent where bold ${\bf 1}$ is used to denote an $unknown$ value. A
calculation using the geometric means method yields $5^{\frac{1}{3}}$, $1$,$%
\frac{1}{5^{\frac{1}{3}}}$ for the weights corresponding to $A$, $B$, and $C$%
. This solution is obviously inaccurate since the exact value for $A/B$ (and
also $B/A$) is given as $1$. On the other hand the discrepancy between $B$
and $C$ can be tolerated even though their ratio was also $1$ since the
ratio was actually assessed as unknown and is denoted here as bold {\bf $1$}
for presentation purposes only.

The simple solution to the above problem is to leave the unknown values as
null entries (e.g., blanks or ``?'' for display purposes) and substitute
estimated values for them when necessary. This can be done (and as later
argued, should be done) by minimizing the global inconsistency. In the above
example bold ${\bf 1}$ can be replaced by $5$ (since the given $5$ divided
by the given $1$ is $5$) in the second row and by $1/5$ (the reciprocal
value) in the third row. This temporary substitution produces the
(unnormalized) weights $\sqrt[3]{5}$, $\sqrt[3]{5}$, $\frac{1}{\sqrt[3]{25}}$
which are more reasonable since now the weights corresponding to $A$ and $B$
are equal as implied by the given ratio of $1$ in the matrix.

\section{Recoverability of missing values}

The pairwise comparisons method is quite flexible as far as recovery of the
unknown values is concerned. On the assumption of the consistency condition $%
a_{ij}\cdot a_{jk}=a_{ik}$, it is possible to recover the entire $n\times n$
matrix from just $n-1$ given elements. They must, however, be present in
specific locations. For example, the given matrix elements can be placed in
the upper row, in the leftmost column, or in the diagonals adjacent to the
main diagonal (that is, the upper diagonal with elements $a_{ij}$ for $i=j-1$
and $2\leq j\leq n$ or the lower diagonal with elements $a_{ij}$ for $i=j+1$
and $1\leq j\leq n-1$).

A different attempt at managing the missing values is presented in \cite
{Harker87}. According to \cite{Harker87}, the purpose of that paper was to
present a method for reducing the number of pairwise comparisons. Harker
proposed a procedure for guiding a decision maker in selecting the
appropriate (i.e., the most important) assessments and recommended that the
decision maker stop making assessments when a sufficient number of
assessments had been made. This interesting approach is not an acceptable
practice for the consistency-driven approach. A decision maker should be
allowed to enter any assessments he/she deems appropriate and the method
should provide a tool for locating the most inconsistent assessments. In the
case of undecidable assessments, however, it is better to leave them void
rather than to force the decision maker to just ''fill-in-the-blanks''.

The proposed procedures for recovering missing values are based on
minimizing a global inconsistency. Missing values should be replaced by
values at which the global inconsistency attains its minimum. Solving this
nonlinear optimization problem is not easy in general and is beyond the
scope of this presentation. However, from the consistency-driven point of
view, the exact solution is not indispensable since the emphasis in this
approach is on a dynamic process involving gradual improvements rather than
finding the exact solution for statically given assessments. In other words,
the journey (to the best assessments) is the reward as long as we make
progress in the right direction (that is reduce the global inconsistency).
Therefore the following suboptimal procedure can be considered: replace the
missing values by the products of the geometric means of the non-null
entries in the respective rows and columns as indicated in the following
formula

\begin{equation}
a_{ij}^{\prime }=GM_{R_{i}}*GM_{C_{j}}= \sqrt[n_{i}]{%
\prod_{k=1}^{n_{i}}a_{ik}} \sqrt[n_{j}]{\prod_{k=1}^{n_{j}}a_{kj}}
\end{equation}

\noindent where the calculation of the geometric means of the $i-$th row ($%
GM_{R_{i}}$) and the $j-$th column ($GM_{C_{j}}$) involves all $n_{i}$ and $%
n_{j}$ non-null entries in the respective rows and columns.

Formula (3) has its genesis in the computation of a consistent approximation
to a generalized pairwise comparisons matrix by geometric means as shown in 
\cite{KoczOrl98}. Proving the convergence of the above procedure will not be
attempted here but the reader is referred to \cite{HolKocz96} for some
useful hints since there are some obvious similarities. It is worthwhile
noting that the above procedure is fairly general since it is not associated
with any particular definition of inconsistency. For the matrix below,
formula (3) gives $a_{12}^{\prime }=\sqrt[4]{1*2*3*4}* \sqrt[4]{1*\frac{1}{%
2.5}*\frac{1}{2.5}*\frac{1}{3}}=1.\,0637$.

With the triad inconsistency definition, another suboptimal inconsistency
minimization procedure can be defined. It is based on a stepwise replacement
of each null entry (one at a time) by a value which minimizes a global
inconsistency for the triads involved. Let us illustrate a stepwise
substitution for a single unknown value (shown as $?$) for the case $n=5$.

\begin{equation}
A=\left| 
\begin{array}{ccccc}
1 & ? & 2 & 3 & 4 \\ 
? & 1 & 2.5 & 2.5 & 3 \\ 
\frac{1}{2} & \frac{1}{2.5} & 1 & 1.3 & 1.8 \\ 
\frac{1}{3} & \frac{1}{2.5} & \frac{1}{1.3} & 1 & 1.5 \\ 
\frac{1}{4} & \frac{1}{3} & \frac{1}{1.8} & \frac{1}{1.5} & 1
\end{array}
\right|
\end{equation}

\noindent The following three triads (in general, there are $n-2$ such
triads) contain the null entry: $(?,2,2.5)$, $(?,3,2.5)$, $(?,4,3)$.
Therefore the replacement value $x$ is a value which minimizes the function

\begin{equation}
f(x)=\max(ix(x,2,2.5),ix(x,3,2.5),ix(x,4,3))\qquad \mbox{in the interval } [%
\frac{1}{5},5]
\end{equation}

\noindent where

\begin{equation}
ix(x,y,z)=\min (\frac{\left| x-y/z\right| }{x},\frac{\left| y-x*z\right| }{y}%
)
\end{equation}

\noindent is the triad inconsistency index introduced in \cite{Kocz93}. A
standard stepping method yields a minimum value of $0.2254$ at $x=1.0328$.
While the local inconsistencies of the three triads are now $0.2254$, $0.1393
$, and $0.2254$ respectively, the global inconsistency of matrix $A$ is
still $0.3333$ because of another triad, $(a_{23},a_{25},a_{35})$.

In the presence of more than one null entry, the above procedure is applied
to each null entry separately, provided no triad includes more than null
entry. Otherwise the order of computing the replacement values may determine
its value. In this case, based on the principle of minimizing the
inconsistency, a replacement value causing the least local inconsistency
(that is, this null value which after computing its replacement, has the
least inconsistency in these triads which include the replacement value) is
set first. The new replacement value is treated as a regular given value
(for this procedure) and the process is repeated. In the most general case
there may be triads containing all three null entries as in the following
matrix

\begin{equation}
A=\left| 
\begin{array}{llll}
1 & ? & a_{13} & ? \\ 
? & 1 & a_{23} & ? \\ 
a_{31} & a_{32} & 1 & a_{34} \\ 
? & ? & a_{43} & 1
\end{array}
\right|
\end{equation}

\noindent However, there must exist other triads from which we can recover
these missing values. In this case, the null entry in $a_{12}$ can be
recovered from $a_{13}/a_{23}$ and the null entry in $a_{24}$ can be
recovered from $a_{23}*a_{34}$. Then we can recover $a_{14}$ from $a_{24}*$ $%
a_{12}$ or we can chose to recover $a_{14}$ from the product of the original
non-null elements $a_{13}*a_{34}$. In general the sequence in which null
entries are recovered may be important and needs further study.

\section{Conclusions}

The above results imply that consistency analysis is essential for handling
uncertainty in the pairwise comparisons method. In practice minimizing the
inconsistency is more important than searching for the most consistent
approximation to an inconsistent pairwise comparisons matrix. The
consistency-driven approach incorporates the reasonable expectation that an
expert is able to reconsider his/her assessments when the most inconsistent
assessments are identified. Thus the location of inconsistencies is crucial
for finding assessments requiring further refinement. This in turn
contributes to improvements in the accuracy of assessments which is a more
constructive approach than not allowing an expert to make a mistake. ``To
err is human'' and error prevention techniques have limited applicability.
However, pointing to an error committed by an expert is the first and
necessary step toward a possible improvement. 

The dynamic process of consistency analysis is facilitated by software,
which locates and displays the most inconsistent assessments (as implemented
in {\em The Concluder} system which has been released to the public domain
and is available at URL http://www.laurentian.ca/www/math/wkocz/ref.html).

\end{document}